%% file: proceedings_versaci.tex
\documentclass[12pt]{article}
\usepackage{epsfig}

\input econfmacros.tex

\def\Title#1{\begin{center} {\Large {\bf #1} } \end{center}}
       
\begin{document}

\Title{New results from KLOE}

\begin{center}{\large \bf Contribution to the proceedings of HQL06,\\
Munich, October 16th-20th 2006}\end{center}

\bigskip\bigskip


\begin{raggedright}  

{\it KLOE collaboration\footnote{
F.~Ambrosino,
A.~Antonelli,
M.~Antonelli,
C.~Bacci,
P.~Beltrame,
G.~Bencivenni,
S.~Bertolucci,
C.~Bini,
C.~Bloise,
S.~Bocchetta,
V.~Bocci,
F.~Bossi,
D.~Bowring,
P.~Branchini,
R.~Caloi,
P.~Campana,
G.~Capon,
T.~Capussela,
F.~Ceradini,
S.~Chi,
G.~Chiefari,
P.~Ciambrone,
S.~Conetti,
E.~De~Lucia,
A.~De~Santis,
P.~De~Simone,
G.~De~Zorzi,
S.~Dell'Agnello,
A.~Denig,
A.~Di~Domenico,
C.~Di~Donato,
S.~Di~Falco,
B.~Di~Micco,
A.~Doria,
M.~Dreucci,
G.~Felici,
A.~Ferrari,
M.~L.~Ferrer,
G.~Finocchiaro,
S.~Fiore,
C.~Forti,
P.~Franzini,
C.~Gatti,
P.~Gauzzi,
S.~Giovannella,
E.~Gorini,
E.~Graziani,
M.~Incagli,
W.~Kluge,
V.~Kulikov,
F.~Lacava,
G.~Lanfranchi,
J.~Lee-Franzini,
D.~Leone,
M.~Martini,
P.~Massarotti,
W.~Mei,
S.~Meola,
S.~Miscetti,
M.~Moulson,
S.~M\"uller,
F.~Murtas,
M.~Napolitano,
F.~Nguyen,
M.~Palutan,
E.~Pasqualucci,
A.~Passeri,
V.~Patera,
F.~Perfetto,
L.~Pontecorvo,
M.~Primavera,
P.~Santangelo,
E.~Santovetti,
G.~Saracino,
B.~Sciascia,
A.~Sciubba,
F.~Scuri,
I.~Sfiligoi,
T.~Spadaro,
M.~Testa,
L.~Tortora,
P.~Valente,
B.~Valeriani,
G.~Venanzoni,
S.~Veneziano,
A.~Ventura,
R.~Versaci,
G.~Xu.
}\\
presented by Roberto Versaci\index{Versaci. R.}\\
Laboratori Nazionali di Frascati dell'INFN\\
Via Enrico Fermi 40,\\
00044 Frascati (RM), ITALY}
\bigskip\bigskip
\end{raggedright}

\section{Introduction}

The most precise determination of $V_{us}$ comes from semileptonic kaon
decays. 
We have measured with the KLOE detector at DA$\Phi$NE, the Frascati
$\phi$-factory, all the experimental inputs to $V_{us}$ for both neutral and
charged kaons. 
Using our results we extract the value of $V_{us}$ with ~0.9\% fractional
error, which is dominated by the theoretical error on the form factor, 
$f_+(0)$.
A new determination of the ratio $V_{us}/V_{ud}$ is also presented, based
on our precise measurement of the absolute branching ratio for the decay
$K \rightarrow \mu \nu (\gamma)$, combined with lattice results for the
ratio $f_K/f_\pi$.
New results on CPT symmetry and quantum mechanics test have also been
achieved, which are based on the first measurement of the charged asymmetry
for $K_S \rightarrow \pi e \nu$ decay and on interferometry studies using
the $\phi \rightarrow K_L K_S \rightarrow \pi^+ \pi^- \pi^+ \pi^-$.

\section{DA$\Phi$NE and KLOE}
The DA$\Phi$NE\index{DA$Phi$NE} e$^+$e$^-$ collider operates at a total
energy $\sqrt{s}=1020$ MeV, the mass of the
$\phi$(1020)-meson.\index{$\Phi$}  

Since 2001, KLOE\index{KLOE} has collected an integrated luminosity of
about 2.5 fb$^{-1}$.
Results presented below are based on 2001-02 data for about 450~pb$^{-1}$.

The KLOE detector consists of a large cylindrical drift chamber surrounded
by a lead/scintillating-fiber electromagnetic calorimeter.
The drift chamber \cite{bib:dc}, is 4~m in diameter and 3.3~m long.
The momentum resolution is $\sigma(p_{T})/p_{T} \sim 0.4\%$.
Two track vertices are reconstructed with a spatial resolution
of $\sim$ 3 mm.
The calorimeter \cite{bib:emc}, composed of a barrel and two endcaps,
covers 98\% of the solid angle.
Energy and time resolution are $\sigma(E)/E = 5.7\%/\sqrt{E[\mbox{GeV}]}$ and
$\sigma(t) = 57 \mbox{ ps}/ \sqrt{E[\mbox{GeV}]} \oplus 100 \mbox{ ps}$.
A superconducting coil around the detector provides a 0.52~T magnetic
field.

The KLOE trigger \cite{bib:trg}, uses calorimeter and drift chamber
information.
For the present analyses only the calorimeter triggers have been used.
Two energy deposits above threshold, $E>50$ MeV for the barrel and
$E>150$ MeV for the endcaps, have been required.

\section{The tag mechanism}

In the rest frame $\phi$-mesons decay into anti-collinear
$K\bar{K}$ pairs (with branching ratios
$\mbox{BR}(\phi \rightarrow K^+ K^- \simeq~49\%)$ and
$\mbox{BR}(\phi \rightarrow K_S K_L \simeq 34\%)$ \cite{bib:pdg}).  
In the laboratory this remains approximately true because of the small
crossing angle of the e$^+$e$^-$ beams.

The decay products of the $K$ and $\bar K$ define, event by event, two
spatially well 
separated regions called the tag and the signal hemispeheres.
Identified $\bar K (K)$ decays tag a $K(\bar K)$ beam and provide an
absolute count, using the total number of tags as normalization. 
This procedure is a unique feature of a $\phi$-factory and provides the
means for measuring absolute branching ratios. 

Charged kaons\index{Kaons} are tagged using the two body decays 
$K^{\pm}\rightarrow \mu^{\pm} \nu_{\mu}$ and
$K^{\pm}\rightarrow \pi^{\pm} \pi^0$. 
$K_S$ are tagged by $K_L$ interacting in the calorimeter ($K_L$-crash);
$K_L$ are tagged detecting $K_S \rightarrow \pi^+ \pi^-$.
For all of cases it is possible to precisely measure the tagged kaon
momentum from the knowledge of the $\phi$ and the tagging kaon momentum.

\section{$K_L$ physics}

As already stated, a pure sample of $K_L$ mesons is selected by the
identification of $K_S \rightarrow \pi^+ \pi^-$ decays.
$K_L$ can either decay in the detector volume
or interact in the calorimeter or escape the detector.

\subsubsection*{\mathversion{bold} Branching ratios of $K_L$ main decays}
Starting from this sample, the $K_L$ branching ratios are evaluated by
counting the number of decays to each channel in the fiducial volume FV and
correcting for the geometrical acceptance, the reconstruction efficiency
and the background contamination.

$K_L$ decays in charged particles are identified by selecting a decay
vertex within the FV along the expected $K_L$ flight direction, as
defined by the tag.
In order to discriminate among the different $K_L$ charged modes the
variable $\Delta_{\mu \pi} = |p_{miss} - E_{miss}|$ is used, where
$p_{miss}$ and $E_{miss}$ are the missing momentum and the missing energy
at the $K_L$ decay vertex, evaluated by assigning to one track the pion
mass and to the other one the muon mass.
Signal counting is thus achieved by fitting the $\Delta_{\mu \pi}$ spectrum
with a linear combination of four Monte Carlo shapes 
($K_L \rightarrow \pi e \nu_e$, $K_L \rightarrow \pi \mu \nu_\mu$, 
$K_L \rightarrow \pi^+ \pi^- \pi^0$, $K_L \rightarrow \pi^+ \pi^-$).

To count $K_L \rightarrow \pi^0 \pi^0 \pi^0$ events, we exploit the time of
flight capability of the calorimeter to reconstruct the neutral vertex
position. 
Such a vertex is assumed to be along the $K_L$ line of flight.
The arrival time of each photon detected in the calorimeter is thus used to 
give an independent determination of $L_K$, the path length of the $K_L$.
Its final value is obtained from a weighted average of the different
measurements.
This decay has been used also to measure the $K_L$ lifetime, 
$\tau_K = 50.92 \pm 0.17 \pm 0.25$ ns, from a fit to the proper time
distribution of neutral decay vertexes \cite{bib:kloe-kl-lifetime}.

Since the geometrical efficiency of the FV depends on $\tau_K$, 
the branching ratios measured by KLOE have been renormalized by imposing
their sum plus the remaining ones ($\approx 0.86\%$ from PDG) to be equal
to one.
This removes the uncertainty due to $\tau_K$, while giving at the same time
a precise determination of the $K_L$ lifetime itself.
The measured branching ratios are \cite{bib:kloe-kl-mainbr}:
\beqa
\mbox{BR}(K_L \rightarrow \pi e \nu_e(\gamma))       = 0.4007\pm 0.0005\pm 0.0004\pm 0.0014\\
\mbox{BR}(K_L \rightarrow \pi \mu \nu_\mu(\gamma))   = 0.2698\pm 0.0005\pm 0.0004\pm 0.0014\\
\mbox{BR}(K_L \rightarrow \pi^+ \pi^- \pi^0(\gamma)) = 0.1263\pm 0.0004\pm 0.0003\pm 0.0011\\
\mbox{BR}(K_L \rightarrow \pi^0 \pi^0 \pi^0(\gamma)) = 0.1997\pm 0.0003\pm 0.0003\pm 0.0019
\eeqan
The corresponding lifetime is:
$\tau_{K_L} = 50.72\pm 0.11\pm 0.13\pm 0.33$ ns. 
It is in agreement with KLOE's previous measurement. 
The two measurements are uncorrelated and can be averaged:
$\tau_{K_L} = 50.84\pm 0.23$ ns.

\subsubsection*{\mathversion{bold} $K_L \rightarrow \pi e \nu_e$ decay: 
                branching ratio and form factor}
From the $K_L e3$ semileptonic decays it is possible to extract the
shape of the vector form factor $f_+(t)$, since extra terms in the matrix
element depend on the lepton mass.
The form factor is usually parametrized as
\beqa
f_+(t) = f_+(0) \left [ 1 + \lambda'_+ \frac{t}{m^2_{\pi^+}} 
+ \frac{\lambda''_+}{2} \left ( \frac{t}{m^2_{\pi^+}} \right )^2 
+ \dots \right ] 
\eeqan
where $f_+(0)$ is evaluated from theory and $t$ is the $K \rightarrow \pi$
four momentum transfer squared of the lepton pair invariant mass.
The parameters $\lambda$',$\lambda''$ are obtained by fitting the spectrum
of $t/m^2_{\pi^+}$ $K_{e3}$ events.
The fit procedure takes into account the efficiency of the selection cuts,
the resolution effects and the background contamination as a function of
$t$.
We find for a fit to $1 + \lambda'_+ t/m^2_{\pi^+}$ \cite{bib:kloe-kle3-ff}:
\beqa
\lambda_+ = (28.6 \pm 0.5 \pm 0.4) \times 10^{-3}
\eeqan
with $\chi^2 /\mbox{dof} = 330 / 363$ ($P(\chi^2)=0.89$);
for the quadratic term:
\beqa
\lambda'_+  = (25.5 \pm 1.5 \pm 1.0) \times 10^{-3}\\
\lambda''_+ = ( 1.4 \pm 0.7 \pm 0.4) \times 10^{-3}
\eeqan
with $\chi^2 /\mbox{dof} = 325 / 362$ ($P(\chi^2)=0.92$).

We also fit the data using a pole parametrization shape,
$f_+(t)/f_+(0) = M^2_V / (M^2_V -t)$. 
We obtain $M_V = (870 \pm 6 \pm 7)$ MeV ($\chi^2 /\mbox{dof} = 326 / 363$ with 
$P(\chi^2)=0.924$).

\subsubsection*{\mathversion{bold} $K_L \rightarrow \pi^+ \pi^-$}
KLOE has also measured the BR of the $K_L \rightarrow \pi^+ \pi^-$
decay.
This has been done measuring the ratio 
$R = \mbox{BR}(K_L \rightarrow \pi^+ \pi^- (\gamma))/ 
     \mbox{BR}(K_L \rightarrow \pi \mu \nu_\mu (\gamma))$
and taking the value of the semileptonic branching ratio previously
measured, since the tagging efficiency are very similar.
The number of events has been obtained fitting the spectrum of the quantity
$\sqrt{E^2_{miss} + p^2_{miss}}$ with a linear combination of the Monte
Carlo shapes for signal and backgrounds corrected for the data/Monte Carlo
ratio.
Thus the result obtained is \cite{bib:kloe-kl-pipi}:
\beqa
\frac{\mbox{BR}(K_L \rightarrow \pi^+ \pi^- (\gamma)}
     {\mbox{BR}(K_L \rightarrow \pi \mu \nu_\mu (\gamma))}
= (0.7275 \pm 0.0042 \pm 0.0054) \times 10^{-2}
\eeqan
using the BR from the semileptonic decay:
\beqa
\mbox{BR}(K_L \rightarrow \pi^+ \pi^- (\gamma)) = (1.963 \pm 0.0012 \pm 0.0017) \times 10^{-3}
\eeqan
This measurement, together with the measurements of the
$\mbox{BR}(K_S \rightarrow \pi^+ \pi^-)$, the $K_L$ and $K_S$ lifetimes, can
be used to determine $|\eta_{+-}|$ and $|\epsilon|$: 
\beqa
|\eta_{+-}| = (2.219 \pm 0.013) \times 10^{-3}\\
|\epsilon|  = (2.216 \pm 0.013) \times 10^{-3}
\eeqan
where for $|\epsilon|$ we have used the world average for 
$Re(\epsilon'/\epsilon)=(1.67\pm0.26)\times10^{-3}$ and assumed
$\mbox{arg}\ \epsilon' = \mbox{arg}\ \epsilon$.

\section{\mathversion{bold} $K_S$ decays}

As already stated, a pure sample of $K_S$ is selected by the
detection of a $K_L$ interaction in the calorimeter ($K_L$-crash).

\subsubsection*{\mathversion{bold} 
$R_S^\pi = BR(K_S \rightarrow \pi^+ \pi^- (\gamma)) / 
           BR(K_S \rightarrow \pi^0 \pi^0)$}
The ratio $R^\pi_S $ is a fundamental parameter of the $K_S$ meson. 
It enters into the double ratio that quantifies direct CP violation in 
$K \rightarrow \pi \pi$ transitions: 
$R^\pi_S / R^\pi_L = 1 - 6 \Re{(\epsilon'/\epsilon)}$.
The most precise measurement was performed by KLOE using data collected in
2000 for an integrated luminosity of ~17 pb$^{-1}$:
$R^\pi_S = 2.236 \pm 0.003 \pm 0.015$ \cite{bib:kloe-ks-piratio2002}.
This result was limited by systematic uncertainties.
A new measurement has been performed using 410 pb$^{-1}$ data collected in
2001 and 2002, inproving on the total error by a factor three.
The $K_S$ decays into two neutral pions are selected by requiring the
presence of at least three EMC clusters with a timing compatible with the
hypothesis of being due to prompt photons (within 5 $\sigma's$) and energy
larger than 20 MeV. 
The selection of charged decays requires for two oppositely charged tracks
coming from the IP. 
The result obtained is:
$R^\pi_S = 2.2555 \pm 0.0056$ \cite{bib:kloe-ks-piratio}.
This result can be compared and averaged with the old one; weighting each
by its independent errors and calculating the average systematic error with
the same weigths gives \cite{bib:kloe-ks-piratio}:
\beqa
R^\pi_S = 2.2549 \pm 0.0054
\eeqan
The result can be combined with the KLOE measurement of
$\Gamma(K_S\rightarrow\pi^{\mp}e^{\pm}\nu(\bar{\nu}) / 
\Gamma(K_S\rightarrow\pi^+\pi^-(\gamma))$ 
to extract the dominant $K_S$ BRs.
For the $\pi\pi$ mode we find \cite{bib:kloe-ks-piratio}:
\beqa
\mbox{BR}(K_S\rightarrow\pi^+\pi^-(\gamma)) = ( 69.196 \pm 0.051) \times 10^{-2} \\
\mbox{BR}(K_S\rightarrow\pi^0\pi^0) = ( 30.687 \pm 0.051) \times 10^{-2}
\eeqan

\subsubsection*{\mathversion{bold} 
$\mbox{BR}(K_S \rightarrow \pi e \nu)$ and charge asymmetry}
The measurement of the BR is an improvement (factor 4 on the total error)
of KLOE's previous result \cite{bib:kloe-kse32002}.
It has been obtained by measuring the ratio 
$\mbox{BR}(K_S \rightarrow \pi e \nu (\gamma)) / 
 \mbox{BR}(K_S \rightarrow \pi^+ \pi^- (\gamma))$ and using
the KLOE's BR for the two bodies decay as normalization.
The event counting is performed by fitting the
$E_{miss} - p_{miss}$ spectrum with a combination of MC shapes for signal
and background \cite{bib:kloe-kse3}:
\beqa
\mbox{BR}(K_S\rightarrow\pi^- e^+ \nu) = ( 3.528 \pm 0.062) \times 10^{-4} \\
\mbox{BR}(K_S\rightarrow\pi^+ e^- \nu) = ( 3.517 \pm 0.058) \times 10^{-4} \\
\mbox{BR}(K_S\rightarrow\pi e \nu)     = ( 7.046 \pm 0.091) \times 10^{-4}
\eeqan
Fitting the ratio of data and MC $t/m^2_{\pi^+}$ distributions we have
measured the form factor slope.
The fit has been performed using only a linear parametrization, since the
available statistics does not allow to be sensitive to a quadratic one.
The result obtained is in agreement with the corresponding value for the
linear slope of the semileptonic $K_L$ form factor.
The slope obtained is $\lambda_+ = (33.9 \pm 4.1) \times 10^{-3}$.
The charge asymmetry measured is \cite{bib:kloe-kse3}:
\beqa
A_S = \frac
{\Gamma(K_S\rightarrow\pi^- e^+ \nu) - \Gamma(K_S\rightarrow\pi^+ e^- \nu)}
{\Gamma(K_S\rightarrow\pi^- e^+ \nu) + \Gamma(K_S\rightarrow\pi^+ e^- \nu)}
= (1.5 \pm 9.6 \pm 2.9) \times 10^{-3} .
\eeqan
The comparison of $A_S$ with the corresponding for $K_L$ allows precision
tests of $CP$ and $CPT$ symmetries.
The difference between the charge asymmetries 
$A_S - A_L = 4 ( Re\ \delta + Re\ x_- )$ signals $CPT$ violation either in
the mass matrix ($\delta$ term) or in the decay amplitudes with
$\Delta S \neq \Delta Q$ ($x_-$ term).
The sum of the asymmetries $A_S + A_L = 4 ( Re\ \epsilon + Re\ y )$ 
is related to $CP$ violation in the mass matrix ($\epsilon$ term) and to $CPT$
violation in the decay amplitude ($y$ term).
$K_S$ and $K_L$ decay amplitudes allow test of the $\Delta S = \Delta Q$
rule through the quantity:
\beqa
Re\ x_+ = \frac{1}{2} \frac
{\Gamma(K_S\rightarrow\pi e \nu) - \Gamma(K_L\rightarrow\pi e \nu)}
{\Gamma(K_S\rightarrow\pi e \nu) + \Gamma(K_L\rightarrow\pi e \nu)} .
\eeqan
The results obtained (using other quantities when needed either from KLOE
when available or from PDG) are:
\beqa
Re\ x_+ = (-0.5 \pm 3.6) \times 10^{-3} \\
Re\ x_- = (-0.8 \pm 2.5) \times 10^{-3} \\
Re\ y   = ( 0.4 \pm 2.5) \times 10^{-3}
\eeqan
they are all compatible with zero.
KLOE has a disposal of a statistic five time bigger, using all the data
available the uncertainty on $A_S$ can be reduced by more than a factor 5.

\subsubsection*{\mathversion{bold} $\mbox{BR}(K_S \rightarrow \pi^0 \pi^0 \pi^0)$}
This decay is a pure $CP$ violating process. 
The related $CP$ violation parameter $\eta_{000}$ is defined as the ratio of
decay amplitudes: 
$|\eta_{000}| = A(K_S \rightarrow 3\pi^0) / A(K_L \rightarrow 3\pi^0) =
\epsilon + \epsilon'_{000}$
where $\epsilon$ describes the $CP$ violation in the mixing matrix and
$\epsilon'_{000}$ is a direct $CP$ violating term.
The signal selection requires six neutral clusters coming from the
interaction point.
Background coming from $K_S \rightarrow \pi^0 \pi^0$ + fake $\gamma$ is
rejected applying a kinematic fit imposing as constraints the $K_S$ mass,
the $K_L$ four momentum and $\beta = 1$ for each photon.
Two pseudo $\chi^2$ variables are then built, $\zeta_3$ which is
based on the best 6 $\gamma$ combination into 3 $\pi^0$ and
$\zeta_2$ which select four out of six $\gamma$ providing the best
agreement with the $K_S \rightarrow \pi^0 \pi^0$ decay.
Events with two charged tracks coming from the interaction point are vetoed.
Using the $K_S \rightarrow \pi^0 \pi^0$ branching ratio as normalization
sample we obtained a 90\% C.L. upper limit \cite{bib:kloe-ks-3pi0}:
\beqa
\mbox{BR}(K_S \rightarrow \pi^0 \pi^0 \pi^0) < 1.2 \times 10^{-7}.
\eeqan
The corresponding 90\% C.L. upper limit on $\eta_{000}$ is:
\beqa
|\eta_{000}| = 
\frac{|A(K_S \rightarrow 3\pi^0)|}{|A(K_L \rightarrow 3\pi^0)|} < 0.018.
\eeqan

\section{Quantum interference in kaons}
KLOE at a $\phi$-factory has the unique possibility for testing QM
and $CPT$ simmetry studying interference in the 
$\phi \rightarrow K_L K_S \rightarrow \pi^+ \pi^- \pi^+ \pi^-$ channel.
Deviation from QM can be parametrized introducing a decoherence parameter
$\zeta$ in the formula for the decay intensity 
\cite{Eberhard:1994pq,Bertlmann:1999np}:
\beqa
I(|\Delta t|) \propto e^{-|\Delta t|\Gamma_L} + e^{-|\Delta t|\Gamma_S} 
-2(1-\zeta)cos(\Delta m|\Delta t|) e^{\frac{\Gamma_S + \Gamma_L}{2}|\Delta t|}
\eeqan
The meaning and value of $\zeta$ depends on the basis used for the initial
state (i.e. $\zeta_{SL}$ for $K_S K_L$ and 
$\zeta_{0\bar{0}}$ for $K^0 \bar{K^0}$).
The results have been obtained performing a fit of the $\Delta t$
distribution. For the decoherence parameter we find
\cite{bib:kloe-interferometry}: 
\beqa
\zeta_{SL}     =& 0.018 \pm 0.040 \pm 0.007               &\qquad \chi^2/\mbox{dof} = 29.7/32 \\
\zeta_{0\bar0} =&(0.10  \pm 0.21  \pm 0.04) \times 10^{-5}&\qquad \chi^2/\mbox{dof} = 29.6/32 .
\eeqan
The results are consistent with zero, therefore there is not evidence for QM
violation. \\
Space-time fluctuactions at the Planck
scale might induce a pure state to become mixed \cite{Hawking:1982dj}.
This results in QM and $CPT$ violation, changing therefore the decay time
distribution of the $K^0\bar{K^0}$ pair from $\phi$ decays.
In some theoretical framework this violation can be
parametrized with the quantities $\gamma$ \cite{Ellis:1983jz}
or $\omega$ \cite{Bernabeu:2003ym}.
Again the values obtained are compatible with zero.
There is no evidence for QM violation \cite{bib:kloe-interferometry}: 
\beqa
\gamma      =& (1.3^{+2.8}_{-2.4} \pm 0.4) \times 10^{-21} GeV 
            &\qquad \chi^2/dof = 33/32 \\
\Re{\omega} =& (1.1^{+8.7}_{-5.7} \pm 0.9) \times 10^{-4}
            &\qquad \chi^2/dof = 29/31 \\
\Im{\omega} =& (3.4^{+4.8}_{-5.0} \pm 0.6) \times 10^{-4} .&
\eeqan

Another test of $CPT$ invariance can be performed via the Bell-Steinberger
relation (BSR) \cite{Bell:1996mn}:
\beqa
\left ( 
   \frac{\Gamma_S + \Gamma_L}{\Gamma_S - \Gamma_L} +i\ \mbox{tan} \phi_{SW} 
\right )
\left ( 
   \frac{Re(\epsilon)}{1 + |\epsilon|^2} -i\ Im(\delta)
\right )
=
\frac{1}{\Gamma_S - \Gamma_L} \sum_f A_L(f)A_S^*(f)
\eeqan
where $\phi_{SW} = \mbox{arctan}(2(m_L-m_S)/(\Gamma_S-\Gamma_L))$.
The Bell-Steinberger relation links a possible violation of $CPT$ invariance 
($m_{K^0} \neq m_{\bar{K^0}}$ or $\Gamma_{K^0} \neq \Gamma_{\bar{K^0}}$) in
the time evolution of the $K^0\bar{K^0}$ system to the observable $CP$
violating interference of $K_L$ and $K_S$ decays into the same final state
$f$.
Any evidence for a non vanishing $Im(\delta)$ can only be due to violation
of: 
  i) $CPT$ invariance; 
 ii) unitarity; 
iii) the time independence of $M$ and $\Gamma$ in the equation which
describes the time evolution of the neutral kaon system within the
Wigner-Weisskopf approximation:
\beqa
i \frac{\partial}{\partial t} \Psi(t) = H\Psi(t) =
\left ( M -\frac{i}{2}\Gamma \right ) \Psi(t) ,
\eeqan
where $M$ and $\Gamma$ are $2 \times 2$ time-independent Hermitian matrices
and $\Psi(t)$ is a two-component state vector in the $K^0-\bar{K^0}$ space.
The result we have obtained (using all experimental inputs from KLOE where
available) are \cite{bib:kloe-bs-relation}:
\beqa
Re(\epsilon)= (159.6 \pm 1.3) \times 10^{-5} \\
Im(\delta)  = (0.4 \pm 2.1)  \times 10^{-5} .
\eeqan
The limits on $Im(\delta)$ and $Re(\delta)$ can be used to constrain the
mass and width difference between the neutral kaons via the relation:
\beqa
\delta= \frac
{i(m_{K^0}-m_{\bar{K}^0})+\frac{1}{2}(\Gamma_{K^0}-\Gamma_{\bar{K}^0})} 
{\Gamma_S-\Gamma_L}
\mbox{cos } \phi_{SW} e^{i\phi_{SW}}[1+O(\epsilon)].
\eeqan
In the limit $\Gamma_{K^0} = \Gamma_{\bar{K}^0}$ 
(i.e. neglecting $CPT$-violating effects in the decay amplitudes)
we obtain the following
bound at 95\% C.L. on the mass difference \cite{bib:kloe-bs-relation}:
\beqa
-5.3\times10^{-19}\ GeV < m_{K^0}-m_{\bar{K}^0} < 6.3\times10^{-19}\ GeV
\eeqan
\begin{figure}[htb]
\begin{center}
\epsfig{file=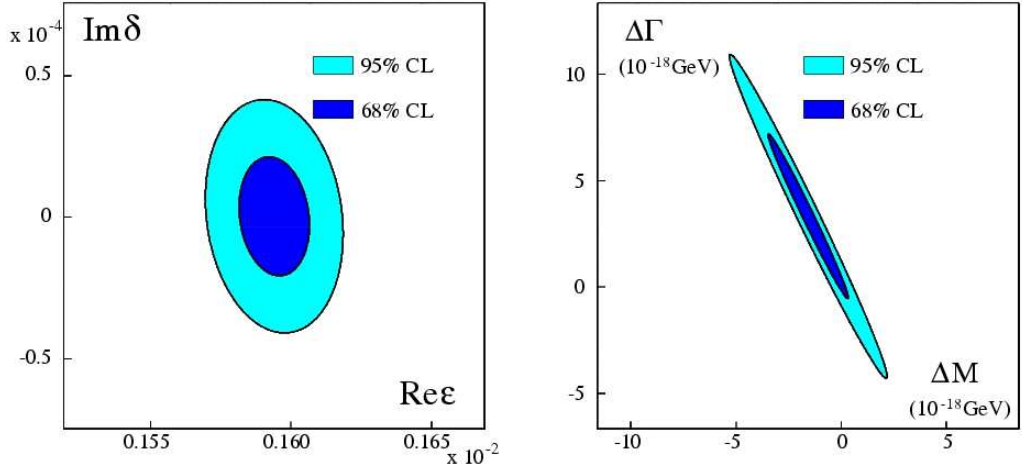,height=2.5in}
\caption{Left: allowed region at 68\% and 95\% CL in the Re($\epsilon$),
         Im($\delta$) plane.
         Right: allowed region at 68\% and 95\% CL in $\Delta M$, 
	 $\Delta \Gamma$ plane.}
\label{fig:vus}
\end{center}
\end{figure}

\section{Charged kaons decays}

As already stated, a pure sample of $K^\pm$ is selected by the
identification of a $K^\mp$ two bodies decay in the drift chamber.

\subsubsection*{Charged kaon lifetime}
Together with the branching ratios of the semileptonic decays, the lifetime
is one of the fundamental experimental inputs for the evaluation of $V_{us}$.
There are two methods available for the measurement: the kaon decay length
and the kaon decay time. 
The two methods allow cross checks and studies of systematics; their
resolutions are comparable.
The first requires a kaon decay vertex in the fiducial volume, then the
kaon is extrapolated backward to the IP taking into account the $dE/dx$ to
evaluate its velocity.
The proper time can be obtained fitting the distribution of
\beqa
\tau^* = \sum_i \Delta T_i 
= \sum_i \frac{\sqrt{1-\beta^2_i}}{\beta_i c}\Delta l_i
\eeqan
The preliminary result we have obtained for the $K^+$ is:
\beqa
\tau^+ = (12.377 \pm 0.044 \pm 0.065) ns
\eeqan
with $\chi^2/\mbox{dof} = 17.7/15$.
The analysis with the second method is still in progress.

\subsubsection*{Branching ratio of the charged kaon semileptonic decays}
The BRs for the two semileptonic decays are obtained performing a fit of
the mass squared of the charged secondary decay product ($m^2_{lept}$),
using the MC distributions for signal and background.
The mass is obtaind via a TOF measurement.
Background from $\mu \nu_\mu$ decay is rejected applying a cut on the
momentum of the charged secondary in the decaying kaon rest frame.
The BRs have been evaluated separately for each tag sample and each charge;
corrections have been applied in order to account for data-MC differences.
The preliminary branching ratios obtained are:
\beqa
\mbox{BR}(K^\pm \rightarrow \pi^0 e^\pm \nu_e (\gamma)
= (5.047 \pm 0.019 \pm 0.039) \times 10^{-2} \\
\mbox{BR}(K^\pm \rightarrow \pi^0 \mu^\pm \nu_\mu (\gamma)
= (3.310 \pm 0.016 \pm 0.045) \times 10^{-2} .
\eeqan

\subsubsection*{\mathversion{bold} $\mbox{BR}(K \rightarrow \mu \nu_\mu (\gamma))$}
The number of signal events has been obtained performing a fit of the 
momentum of the charged secondary in the decaying kaon rest frame.
Background has been identified as any event having a $\pi^0$ in the final
state. 
The efficiency has been evaluated directly on data using a sample selected
only with calorimeter informations.
The result obtained is \cite{bib:kloe-kmu2}:
\beqa
\mbox{BR}(K^+ \rightarrow \mu^+ \nu_\mu (\gamma)) = 0.6366 \pm 0.0009 \pm 0.0015
\eeqan

\section{$V_{us}$ summary}
The KLOE results on semileptonic decays on both neutral and charged
kaons, can be used together with results from other experiments in order to
evaluate $V_{us}$ and check the unitarity of the first row of the CKM
matrix.\\
Averaging over all the available experimental inputs according to the
procedure specified in \cite{Mescia:2004xd}, it is possible to extract the
world average: 
\beqa
V_{us} \times f_+(0) = 0.2164 \pm 0.0004
\eeqan
which can be compared with the value expected from unitarity of CKM matrix
using $V_{ud}$ from \cite{Marciano:2005ec}:
\beqa
V_{us} \times f_+(0) = 0.2187 \pm 0.0022 .
\eeqan
We use $f_+(0) = 0.961 \pm 0.008$, computed by Leutwyler and Roos
\cite{Leutwyler:1984je}.
\begin{figure}[htb]
\begin{center}
\epsfig{file=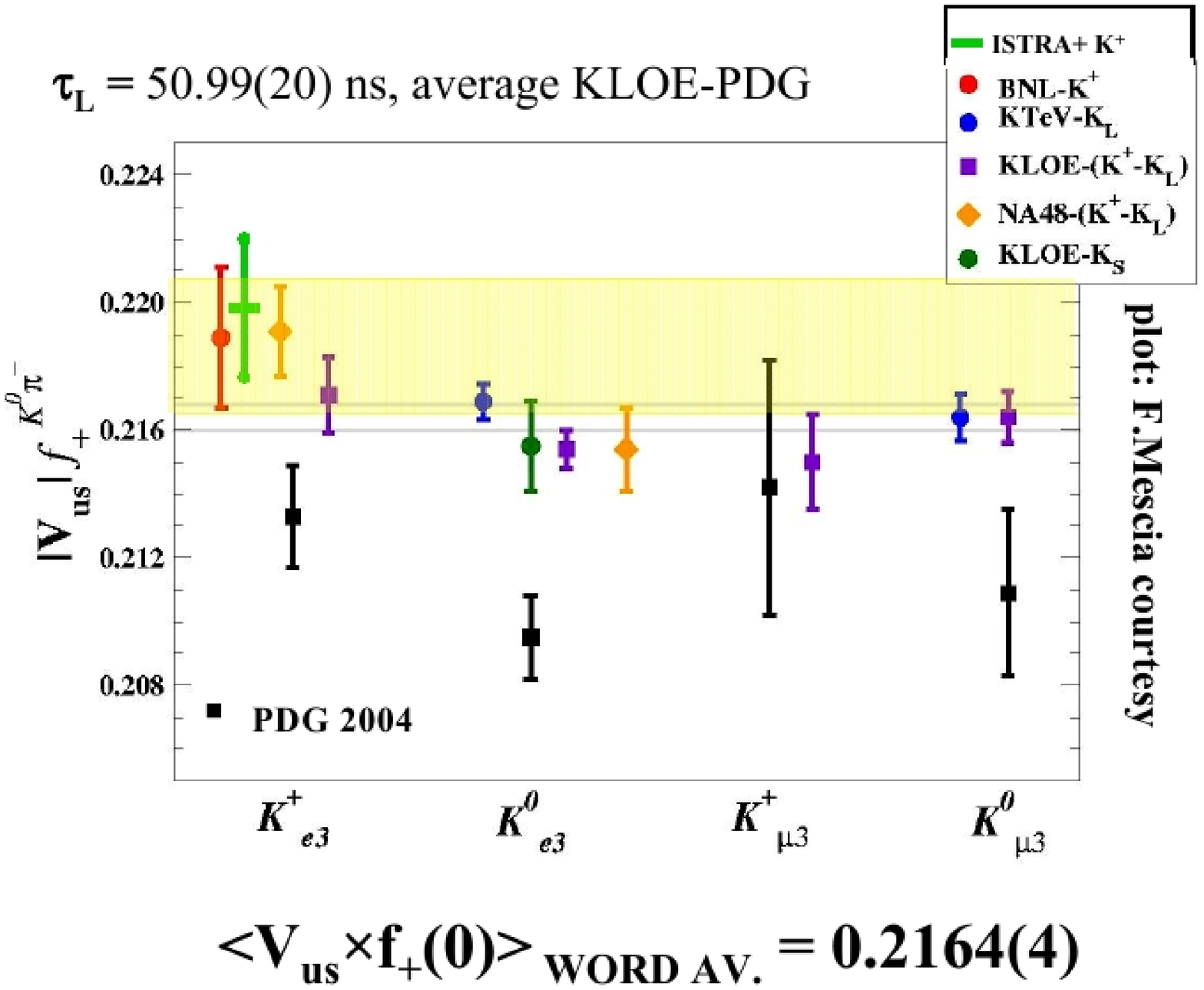,height=2.5in}
\epsfig{file=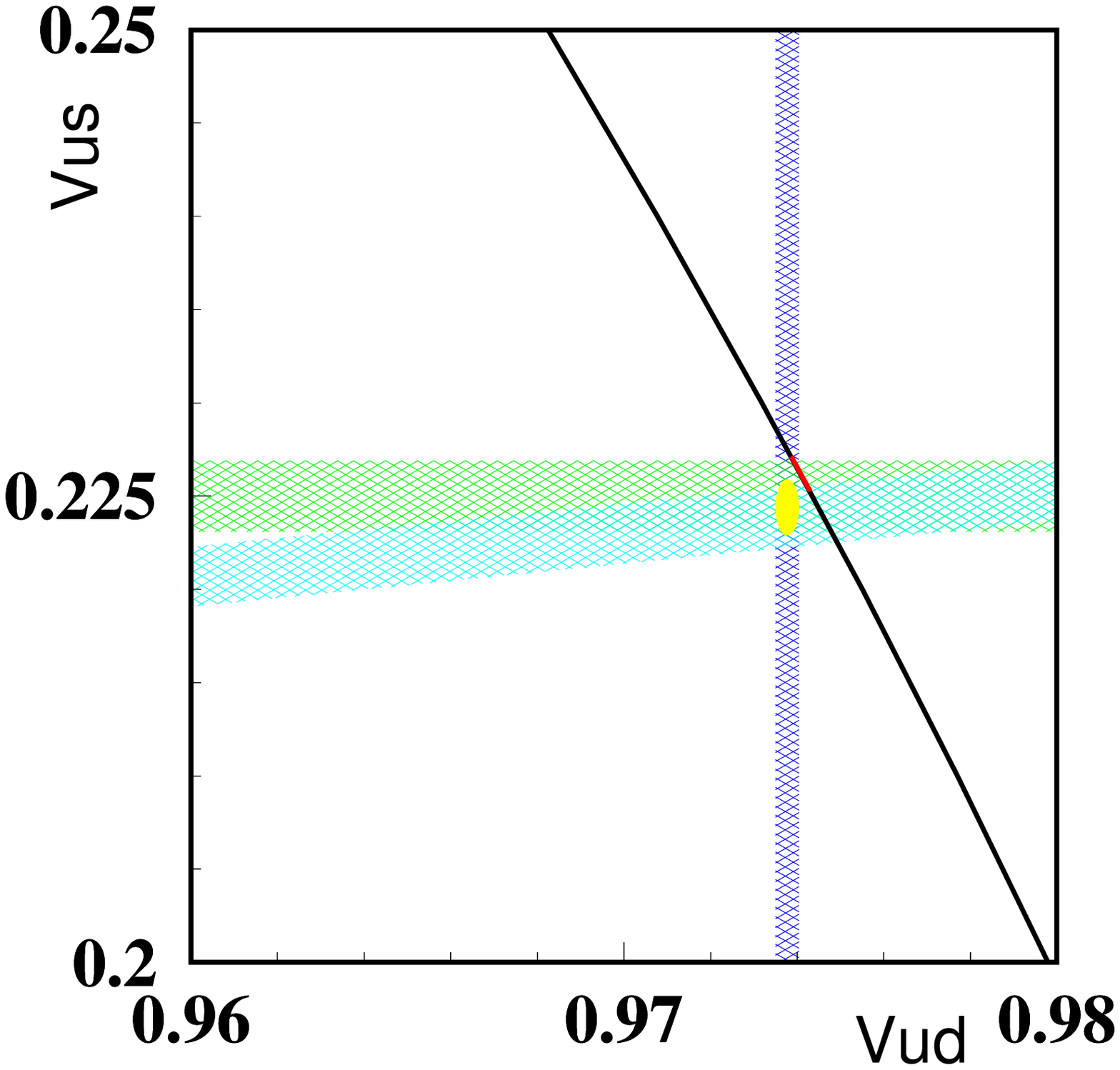,height=2.5in}
\caption{Left: $V_{us} \times f_+(0)$ world average. 
         Right: $V_{us}-V_{ud}$ plane.}
\label{fig:vus}
\end{center}
\end{figure}
It is also possible to use the charged kaon leptonic decay to evaluate
$V_{us}$ using lattice calculation of the ratio $f_K/f_\pi$ as pointed out
in \cite{Marciano:2004uf}.
Using the latest value from the MILC collaboration for the ratio of the
decay constants \cite{Bernard:2006wx} we find:
\beqa
\frac{V_{us}}{V_{ud}} = 0.2286^{+0.0020}_{-0.0011}.
\eeqan
This value can be fitted together with $V_{us}$ from kaon semileptonic decays
and $V_{ud}$ from nuclear beta decays, obtaining:
\beqa
&V_{us} &= 0.2246^{+0.0009}_{-0.0013}\\
&V_{ud} &= 0.97377 \pm 0.00027
\eeqan
with $\chi^2/dof = 0.046/2$, P($\chi^2$) = 0.97. 
Imposing also the unitarity constraint 
(see the right panel of figure \ref{fig:vus}):
\beqa
&V_{us} &= 0.2257 \pm 0.0007\\
&V_{ud} &= 0.97420 \pm 0.00016
\eeqan
with $\chi^2/dof = 3.94/1$, P($\chi^2$)  = 0.05.

%

\end{document}

%% file: econfmacros.tex
\def\beq{\begin{equation}}
\def\eeq#1{\label{#1}\end{equation}}
\def\eeqn{\end{equation}}

\def\beqa{\begin{eqnarray}}
\def\eeqa#1{\label{#1}\end{eqnarray}}
\def\eeqan{\end{eqnarray}}

\let\bar=\overbar

\def\Dslash{\not{\hbox{\kern-4pt $D$}}}
\def\dslash{\not{\hbox{\kern-2pt $\del$}}}

\def\msb{{\bar{\ssstyle M \kern -1pt S}}}

